\documentclass[%
prl, twocolumn,
superscriptaddress,
 amsmath,amssymb,aps,
]{revtex4-2}

\usepackage{graphicx}% Include figure files
\usepackage{dcolumn}% Align table columns on decimal point
\usepackage{bm}% bold math
\usepackage{xcolor}
\usepackage{color,soul} % highlight
\usepackage{physics}
\usepackage[normalem]{ulem}

\usepackage[colorlinks=true, pdfstartview=FitV, linkcolor=blue, citecolor=blue, urlcolor=blue]{hyperref}
\usepackage{amsmath}
\usepackage{amssymb}
\usepackage{natbib}
\usepackage{bbold}
\usepackage{multirow}
\usepackage{booktabs}

\newcommand{\Ha}[1][]{\hat{\mathcal{H}}_{#1}}

\newcommand{\ca}[1][]{\hat{a}_{#1}}
\newcommand{\cad}[1][]{\hat{a}_{#1}^\dag}
\newcommand{\cc}[1][]{\hat{c}_{#1}}
\newcommand{\ccd}[1][]{\hat{c}_{#1}^\dag}

\newcommand{\Dc}{\Delta_c}
\newcommand{\wDc}{\widetilde{\Delta}_c}

\newcommand{\kk}[1][]{\vb{k}_{#1} }

\newcommand{\FO}[1][]{\hat{\Psi}_{#1}}
\newcommand{\FOd}[1][]{\hat{\Psi}_{#1}^\dag}
\newcommand{\FOx}[1][\textbf{x}]{\FO(#1)}
\newcommand{\FOdx}[1][\textbf{x}]{\FOd(#1)}

\newcommand{\PR}[1][]{\frac{Vp^2}{U_0}}

\begin{document}

\preprint{APS/123-QED}

\title{Pauli crystal superradiance}% Force line breaks with \\

\author{Daniel Ortuño-Gonzalez}
\affiliation{%
Institute for Theoretical Physics, ETH Z\"urich, Wolfgang-Pauli-Str. 27, CH-8093 Zurich, Switzerland
}%

\author{Rui Lin}
\email{linrui@quantumsc.cn}
\affiliation{Quantum Science Center of Guangdong-Hong Kong-Macao Greater Bay Area, 518045 Shenzhen, China}

\author{Justyna Stefaniak}
\affiliation{Institute for Quantum Electronics, ETH Z\"urich, Otto-Stern-Weg 1, CH-8093 Zurich, Switzerland}
\author{Alexander Baumgärtner}
\affiliation{Institute for Quantum Electronics, ETH Z\"urich, Otto-Stern-Weg 1, CH-8093 Zurich, Switzerland}
\affiliation{JILA, University of Colorado and National Institute of Standards and Technology, Boulder, Colorado 80309, USA}
\author{Gabriele Natale}
\affiliation{Institute for Quantum Electronics, ETH Z\"urich, Otto-Stern-Weg 1, CH-8093 Zurich, Switzerland}

\author{Tobias Donner}%
\email{donner@phys.ethz.ch}
\affiliation{Institute for Quantum Electronics, ETH Z\"urich, Otto-Stern-Weg 1, CH-8093 Zurich, Switzerland}

\author{R. Chitra}%
\affiliation{%
Institute for Theoretical Physics, ETH Z\"urich, Wolfgang-Pauli-Str. 27, CH-8093 Zurich, Switzerland
}%

%  \homepage{http://www.Second.institution.edu/~Charlie.Author}
% \affiliation{
%  Second institution and/or address\\
%  This line break forced% with \\
% }%
% \affiliation{
%  Third institution, the second for Charlie Author
% \affiliation{%
%  Authors' institution and/or address\\
%  This line break forced with \textbackslash\textbackslash
% }%

% \collaboration{CLEO Collaboration}%\noaffiliation

\date{\today}% It is always \today, today,

\begin{abstract}
Pauli crystals are unique geometric structures of non-interacting fermions, resembling crystals, that emerge solely from Fermi statistics and confinement. Unlike genuine quantum crystals that arise from interparticle interactions, Pauli crystals  do not break translation symmetry  but nonetheless exhibit  nontrivial many-body correlations.  In this Letter, we explore Pauli crystal formation in a  cavity-fermion setup.  We analytically show that when coupled to a cavity,  degeneracy in Pauli crystals can  trigger zero-threshold transitions to superradiance.   This superradiance is accompanied by the emergence of a genuine quantum crystalline state, wherein the atomic density is periodically modulated.   We substantiate our findings  using state-of-the-art    numerical simulations.  The combined interplay between statistics, confinement geometry and interactions mediated by light thus facilitates a novel pathway to quantum crystallization.
\end{abstract}

\maketitle

A quantum crystal is a state characterized by  spontaneously broken translation symmetry in free space and strong zero-point fluctuations about its equilibrium position~\cite{guyer1970physics}. 
Interparticle interactions lie at the core of such crystallization phenomena triggered by the competition between two energy scales, associated with distinct ground states of a many-body system. Crystal states arise in both bosonic and fermionic systems and well-known  examples in the condensed matter realm include rare gas crystals~\cite{pollack1964solid}, molecular solids, Wigner crystals~\cite{wigner1934interaction}, Rydberg crystals~\cite{brune2020evaporative, schachenmayer2010dynamical}, or supersolids~\cite{Boninsegni2012Colloquium,choi2010evidence}.  In the past decade, quantum engineered systems have emerged as a platform of choice to explore diverse facets of crystallization~\cite{cazorla2017simulation}.

A complementary pathway to crystallization is provided by \emph{confinement} of long-range interacting quantum systems, as evinced by the emergence of complex states like bilayer ion crystals~\cite{hawaldar2024bilayer},  molecular dipolar crystals in bosonic and fermionic gases~\cite{rabl2007molecular}, and mesoscopic supersolids~\cite{golomedov2011mesoscopic, Tanzi2019Observation, guo2019supersolid,Chomaz2019Longlived} among others. Nonetheless, interactions remain the key to such crystallization.
Curiously, in the opposite limit of zero interactions, strong confinement of fermions leads to the formation of quantum geometric structures known as Pauli crystals~\cite{rakshit2017observability, gajda2020pauli,xiang2023pauli}. They manifest complex symmetries in position and momentum spaces in high-order correlations resembling crystals at ultra-low temperatures, but without spontaneously breaking translational symmetry as real crystals. They are thus a pure manifestation of Fermi statistics in confined geometries, wherein statistics induces many-body correlations. Pauli crystals have recently been observed using a few ultra-cold spin-polarized fermionic $^6$Li atoms confined to a tight harmonic trap~\cite{holten2021observation}. 

\begin{figure}
  \includegraphics[width=1\columnwidth]{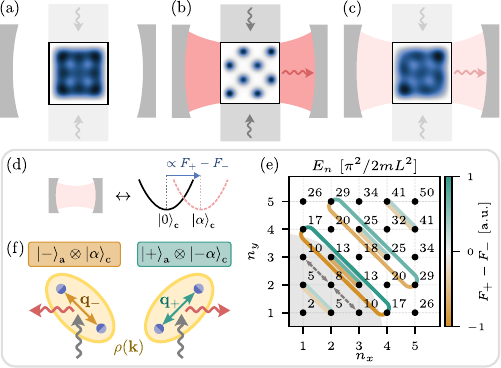}
  \caption{\textbf{(a-c)} Pauli crystals in a $2\lambda \times 2 \lambda$ square box  in an optical cavity and driven by  laser pumps (grey arrows). \textbf{(a)} For $N=8$ fermions and at weak pump strength, the unique Pauli crystal ground state remains unchanged and the cavity is empty. \textbf{(b)} For $N=8$ fermions and beyond a critical pump strength, the Pauli crystal self-organizes into a checkerboard pattern, while triggering superradiant scattering of photons  into the cavity (red arrows). \textbf{(c)} With $N=7$ fermions, the Pauli crystal has degenerate ground states coupled to cavity. The system chooses a configuration maximizing the coherent cavity field already at vanishing pump field. 
  \textbf{(d)} Cavity field as a net displaced harmonic oscillator by the atomic order. \textbf{(e)} Synthetic momentum lattice for a single-particle in a square box, with each eigenmode denoted by its energy. Highlighted pairs of degenerate eigenmodes have cavity mediated coupling  $F_+-F_-= \pm W_{n_a, n_b}^2$ [cf.\ Eqs.\,(\ref{eq:cavityH}-\ref{eq:interference})]. Coupling sign reflects the two $\mathbb{Z}_2$ cavity states associated to $\vb{q}_\pm$ (colored links) with maximum value at unit filling $\nu=1$ (edge of shaded circle). As the spinless fermions fill the eigenmodes, the system exhibits soft transitions if the most energetic fermion occupies an (anti-)symmetric superposition of one of these eigenmode pairs. Otherwise, the system exhibits usual Dicke superradiance between eigenmodes with different energies (dashed arrows). \textbf{(f)} (Anti-)symmetric superposition of Pauli crystals correspond to a squeezed Fermi sea along directions of $\vb{q}_\pm$ which scatters light in the form of a coherent state $\ket{\mp \alpha}$.} 
  
  \label{fig:fig1}
\end{figure}

Understanding how interactions influence Pauli crystals–and how they interplay with correlations arising from quantum statistics–poses a natural extension of current studies.  A promising arena to explore such physics is the hybrid platform of laser-pumped cold gases coupled to high-finesse optical cavities~\cite{Mivehvar2021Cavity}. The tunable cavity-matter coupling induced by the laser with wavelength $\lambda$ creates effective atom-atom interactions that, competing with kinetic energy, favor a $\lambda$-periodic checkerboard modulation of the atomic density over a uniform distribution~\cite{nagy2010dicke}. At a critical coupling, the system undergoes a $\mathbb{Z}_2$ symmetry breaking quantum phase transition to a self-ordered superradiant crystalline state [Fig.~\ref{fig:fig1}(a,b)], which scatters light efficiently into the cavity, rendering it a coherent state. The predicted superradiance and the associated density modulations were observed experimentally for both bosonic~\cite{baumann_2010_dicke,kessler_2014_steering} and fermionic systems~\cite{zhang2021observation, helson2023densitywave}.
    
\begin{figure}
  \includegraphics[width=1\columnwidth]{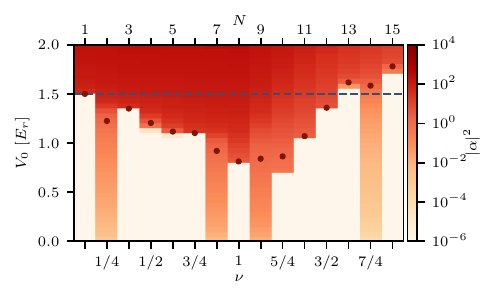} 
  \vspace{-7mm}
  \caption{ Phase diagram showing the cavity field (colorbar in log-scale) as a function of pump strength $V_0$ for Pauli crystals with different filling factors $\nu$. Zero threshold transitions take place when the most energetic fermion of an open-shell Pauli crystal is susceptible to the cavity, as illustrated by the colored links in Fig.~\ref{fig:fig1}(d). For guide of eyes and comparison, dots mark the (arbitrary) $|\alpha|^2 = 1$ threshold, while the dashed line shows the threshold if the system were bosonic. }
  \label{fig:fig2}
\end{figure}

While prior studies focused exclusively on bulk systems and their phase transitions, in this Letter we target the opposite regime of low densities of tightly confined degenerate fermions. We show that Fermi statistics induced correlations inherent to two-dimensional Pauli crystals may dramatically enhance the susceptibility to superradiance via zero-threshold transitions to superradiant quantum crystals [Fig.~\ref{fig:fig1}(c)], termed soft transitions. 
Using a box potential which admits simple analytical treatments while being experimentally accessible~\cite{gaunt2013bose,navon2021quantum}, we present analytical criteria dictating the existence of such soft transitions for different fermion numbers and extents of confinement.  Compared to conventional Dicke superradiance~\cite{dicke54,hepp73,wang73,carmichael73}, these soft transitions take advantage of symmetry breaking within a degenerate subspace accessed through Pauli exclusion~\cite{gajda2020pauli}, and do not need to overcome a kinetic energy barrier.
We consolidate our analytical results with in-depth numerical simulations using high-performance computing methods, which map out the full phase diagram (Fig.~\ref{fig:fig2}) and the associated observables (Fig.~\ref{fig:fig3}). The mechanism we explore is inherently different from Umklapp superradiance, where the momentum transfer required for self-ordering is fine tuned to coincide with the size of the Fermi sphere~\cite{piazza2014Umklapp, Keeling2014Fermionic,
chen2014superradiance} in fermionic systems coupled to a cavity.

We consider $N$  spinless fermions of mass $m$  in a two dimensional square box of side length $L$ coupled to a single-mode cavity with resonance frequency $\omega_c$, wavevector $\mathbf{k}_c$, and maximal single-atom dispersive shift $U_0$, as sketched in Fig.~\ref{fig:fig1}(a). The atoms are driven by two counterpropagating transverse pump laser fields with orthogonal polarizations, frequency $\omega_p$, wavevector $\mathbf{k}_p$, and lattice depth $V_0$. We consider the cavity and pump share the same wavelength $\lambda=2\pi/|\kk[p]|=2\pi/|\kk[c]|$, and 
are directed along the $x$ and $y$ directions, respectively. The coupling to the cavity introduces a recoil process, where a recoil momentum $\pm\vb{q}_\pm = \pm(\vb{k}_p \pm \vb{k}_c)$ is transferred between two atoms~\cite{mottl2012roton,lin2019superfluid}, which by construction fulfill $\vb{q}_+\cdot\vb{q}_- =0$.
In the dispersive regime, the atomic excited state can be adiabatically eliminated, and we obtain the following effective Hamiltonian in the frame rotating at $\omega_p$ ($\hbar=1$):
\begin{align}
  \hspace{-1.4mm}\Ha =& \int \dd\vb{x} \FOdx \bigg[ -\frac{\nabla^2}{2m} + \hat{V}_\text{trap}(\vb{x})+ U_0 \cad\ca \sin^2(\vb{k}_c\vb{x})+ \nonumber \\ 
  & \hspace{-2mm} \sqrt{U_0 V_0}(\ca+\cad)\sin(\vb{k}_p\vb{x})\sin(\vb{k}_c\vb{x})\bigg] \FOx -\Dc \cad\ca. \label{eq:Hamiltonian}
\end{align}
Here, $\ca$ and $\FOx$ denote the annihilation operators for the cavity and atomic fields, satisfying $[\ca,\cad]=1$ and $\{\FOx,\FOdx[\vb{x}']\}=\delta(\vb{x}-\vb{x}')$, respectively; while $\Delta_c=\omega_c-\omega_p$ is the cavity detuning.
In this Letter, we consider a specific implementation of $\beta=2L/\lambda\in2\mathbb{Z}^+$, where the cavity-atom system has a $\mathbb{Z}_2$ symmetry $\hat{a}\to-\hat{a}$, $(x,y)\to(L-x,y)$,  with an associated parity operator $\hat{P}$ which is also reflected between the recoil momenta $\vb{q}_+\to \vb{q}_-$. The square box trap imposes a symmetry of the group $D_4$, which is equipped with a degenerate irreducible representation facilitating the breaking of parity. This manifests as soft transitions and is straightforwardly generalizable to other symmetries.

The atoms are subject to boundary condition $(x,y)\in [0,L]^2$ imposed by the square box potential $\hat{V}_\mathrm{trap}(\mathbf{x})$. This imposes the atomic eigenmode structure $|n_x,n_y\rangle = |n_x\rangle\otimes|n_y\rangle$ where the associated real space wavefunctions are
$\psi_{n_i}(x_i) = \sqrt{\frac{2}{L}} \sin(\frac{n_i \pi x_i}{L} )$
with $n_i\in\mathbb{N}^+,i \in \{x,y\}$, and the energies are $E_{n_x,n_y} = (n_x^2+n_y^2)\pi^2/2mL^2$, see Supplemental Material (SM)~\cite{supmat}. In momentum space, these states center around $(k_x,k_y)=(n_x,n_y)\pi/L$, and thus constitute an effective momentum lattice, as depicted in Fig.~\ref{fig:fig1}(e). In the absence of coupling to the cavity, the fermions fill the $N$ lowest eigenmodes, and the many-body ground state is thus constructed by the corresponding Slater determinant. These states form the geometrical structures of the fermions known as  Pauli crystals~\cite{gajda2020pauli}. As elaborated in the SM, there are two classes of ground states: {\it closed-} and {\it open-shell} configurations depending on whether the ground state's most energetic particle is in a filled (unique ground state) or partially filled energy level (degenerate ground states).  Closed-shell configurations have momentum distributions which are invariant under  $k_i \rightarrow -k_i$ with $i\in \{x,y\}$, whereas  degenerate open-shell configurations host superposition states that violate this symmetry~\cite{supmat}.

Beyond a critical coupling to the cavity, the bulk system undergoes a $\mathbb{Z}_2$ symmetry-breaking Dicke-type transition from a normal to a superradiant phase~\cite{nagy_2010_dicke,baumann_2010_dicke}. In this phase, atoms self-organize into a checkerboard configuration dictated by the recoil processes $\vb{q}_\pm$.
Intriguingly, the Pauli crystal coupled to a cavity may host soft superradiant transitions immediately at infinitesimal cavity-atom couplings. This distinction can be captured by a perturbation analysis. To lowest order in the cavity-atom coupling, the atoms effectively apply a displacement force to the cavity's harmonic oscillator mode, see SM~\cite{supmat},  
 \begin{equation}
 \Ha[c] = -\wDc \cad\ca +  \sqrt{U_0 V_0}(F_+ - F_-)(\ca +\cad), \label{eq:cavityH}
\end{equation}
where $\wDc = \Dc - \mathcal{B} <0$ is an effective cavity detuning defined through bunching parameter $\mathcal{B}$~\cite{supmat}, and $F_\gamma = F(\vb{q}_\gamma)\equiv - \frac{1}{4}\sum_{s\in \pm} \int \dd \vb{k} \ev*{\hat{\Psi}^\dag_{\vb{k}+s \vb{q}_\gamma} \hat{\Psi}_{\vb{k}}} $. The atomic order parameter $F_+-F_- = \int \dd\vb{x} \langle\FOdx\FOx\rangle\sin(\vb{k}_p\vb{x})\sin(\vb{k}_c\vb{x})$
reflects the atomic configuration in the absence of the cavity.

The abundant configurations of Pauli crystals provide an opportunity to spontaneously break the symmetry between $\vb{q}_\pm$. In a {\it closed-shell} Pauli crystal, where the most energetic fermion is in a fully occupied energy level, the ground state is unique. In this case, the ground state inherits the same $\mathbb{Z}_2$ symmetry from the Hamiltonian, and has vanishing atomic order parameter, $F_+=F_-$. Correspondingly, the cavity field is not macroscopically excited $\langle \hat{a}\rangle = 0$. Similar arguments apply for bulk Bose-Einstein condensates and Fermi gases as well, whose configurations are characterized by a uniform distribution. 
However, in an {\it open-shell} crystal, where the most energetic level is partially filled, the ground state is two-fold degenerate (See SM~\cite{supmat}), and the ground state does not necessarily retain the symmetry of the Hamiltonian. This configuration is potentially associated with a finite atomic order parameter $F_+\neq F_-$ through a deformation of the occupied $\vb{k}$-space along the dominant momenta $\vb{q}_\pm$ while keeping its volume, which we term Fermi sea squeezing, triggering a soft transition to superradiance  [Fig.~\ref{fig:fig1}(f)].

To further unveil the interplay between cavity-mediated interactions and statistics in few-body systems, we proceed to the atomic perspective, and rewrite the cavity-atom coupling in the basis spanned by the atomic eigenmodes $\ket{\underline{n}}=\ket{n_x, n_y}$
\begin{equation}
    \hat{\mathcal{H}}_\mathrm{int} = \sqrt{U_0 V_0} (\ca +\cad) \sum_{\underline{n}, \underline{m}}W_{n_x,m_x}W_{n_y,m_y} |\underline{n}\rangle\langle\underline{m}|,  \label{eq:interference} \\ 
%\sum_{\substack{n_x,n_y, \\ m_x,m_y}}
  \end{equation}
with $W_{n_\mu,m_\mu} = \bra{n_\mu} \sin(2 \pi \hat{\mu}/\lambda) \ket{m_\mu}$, $\mu=x,y$, evaluated as $W_{n,m} = 4[1-(-1)^{m+n}]\beta  mn/[\pi(\beta^2-m^2-n^2)^2 - 4\pi m^2 n^2]$. 
The cavity-atom coupling effectively introduces hoppings between atomic eigenmodes on the synthetic momentum lattice~\cite{rosamedina2022observing}. 
The onset of superradiance depends on the fermion most susceptible to this effective coupling. 
In a closed-shell crystal, the couplings most relevant to superradiance are those connecting filled eigenmodes $|n_{1x},n_{1y}\rangle$ with empty ones $|n_{2x},n_{2y}\rangle$. Each eigenmode pair can be regarded as a two-level Dicke model with critical coupling controlled by the energy differences and the effective couplings between them. The smallest critical coupling among all these pairs is thus the critical coupling of the Pauli crystal. Due to the finite energy differences between every pair, the critical coupling is non-zero.

An open-shell crystal, however, provides a more subtle structure because its low-energy atomic dynamics take place predominantly in a subspace spanned by the degenerate eigenmodes $|n_a,n_b\rangle$ and $|n_b,n_a\rangle$, see SM~\cite{supmat}. 
When $n_a+n_b$ is even, $W_{n_a,n_b}=0$. The degenerate eigenmodes destructively interfere with each other through the recoil process, and the cavity does not mediate couplings within the degenerate subspace. 
However, when $n_a+n_b$ is odd and thus $W_{n_a,n_b}\neq0$, the cavity-mediated coupling lifts the degeneracy, and selects the states $(|n_a,n_b\rangle\pm|n_b,n_a\rangle)/\sqrt{2}$ most compatible with its recoil wavevectors $\vb{q}_\pm$ to maximize the atomic order parameter $|F_+-F_-|$ to $W_{n_a,n_b}^2$. 
Consequently, the two-fold degenerate Pauli crystal-cavity states $\ket{1}_{\text{a}}\otimes\ket{0}_{\text{c}}$ and $\ket{2}_{\text{a}}\otimes\ket{0}_{\text{c}}$  where $\ket{1}_{\text{a}}, \ket{2}_{\text{a}} $  and  $\ket{0}_{\text{c}}$ denote the two Pauli crystal configurations and the vacuum cavity state respectively, split into four states: two lowest energy states $\ket{-}_{\text{a}}\otimes\ket{\alpha_0}_{\text{c}}$ and $\ket{+}_{\text{a}}\otimes\ket{-\alpha_0}_{\text{c}}$ and two excited states $\ket{-}_{\text{a}}\otimes\ket{-\alpha_0}_c$ and $\ket{+}_{\text{a}}\otimes\ket{\alpha_0}_{\text{c}}$  where $\ket{\pm}_{\text{a}}$ is the (anti\mbox{-})symmetric superposition of $\ket{1}_{\text{a}}$ and $\ket{2}_{\text{a}}$ and $\ket{\alpha_0}_{\text{c}}$ is a coherent state of the cavity.  Each new ground state is one of the $\mathbb{Z}_2$ configurations with the momentum distribution of the atoms squeezed along $\vb{q}_\pm$, leading to a non-zero expectation value for the cavity. 
In contrast to the nondegenerate eigenmodes with fixed parity $\langle \hat{P}\rangle =\pm1,$ the degeneracy generates states with varying parities $\langle \hat{P}\rangle \in[-1,1]$, which allows the cavity-atom coupling to break the parity maximally with $\langle \pm|\hat{P}|\pm\rangle_a = 0$.

To verify our theory of Pauli-crystallinity-assisted soft superradiant transitions, we perform numerical simulations using the Multiconfigurational Time-Dependent Hartree Method for Indistinguishable Particles (MCTDH-X)~\cite{lode2024ultracold,fasshauer2016multiconfigurational,lode2020mctdh-x,Lin2020MCTDHX,molignini2025many}, see SM~\cite{supmat}. 
MCTDH-X has been successfully applied to analyzing static and dynamic properties of cavity-boson systems~\cite{molignini2018superlattice,lin2020pathway,lin2021mott,rosamedina2022observing}, but its application to two-dimensional fermionic systems~\cite{fasshauer2016multiconfigurational,xiang2023pauli} is far more computationally challenging due to the underlying Slater determinant ansatz.

Specifically, we simulate an experimentally viable setup of fermions in a box of size $L= 2 \lambda$  corresponding to $\beta=4$ with particle number ranging \(N=1,2,\dots,15\). This corresponds to $N$ fermions that occupy a lattice with $ 2L^2/\lambda^2 = \beta^2/2 = 8$ sites once superradiance starts, thus defining filling fraction as $\nu=N/8$ ranging from $1/8$ to $15/8$. The system parameters are inspired by  current experimental setup \(N U_0 = -2\pi \times 5.22\) MHz, \(\Delta_c = -2\pi \times 7.50\) MHz, a wavelength of \(\lambda = 780\) nm corresponding to a recoil energy of \(E_r = 2\pi \times 3.77\) kHz, and a dissipation of \(\kappa = 2\pi \times 147\) kHz \cite{li2021first, dreon2022self}. The pump field is  red-detuned with respect to the atomic transition, i.e. $V_0, U_0<0$.
The cavity is treated as a coherent state $\ca\to \alpha=\langle\ca\rangle$ effectively providing a one-body potential, which follows the master equation  $\partial_t \alpha = i\langle [\Ha,\ca] \rangle-\kappa\alpha$.
We simulate the steady states for different values of $V_0$ between 0 and \(2E_r\). As summarized in Fig.~{\ref{fig:fig2}, we observe that the Pauli pressure significantly impacts the superradiance threshold, showing both soft and hard transitions.

\begin{figure}[t]
\vspace{10pt}
  \includegraphics[width=1\columnwidth]{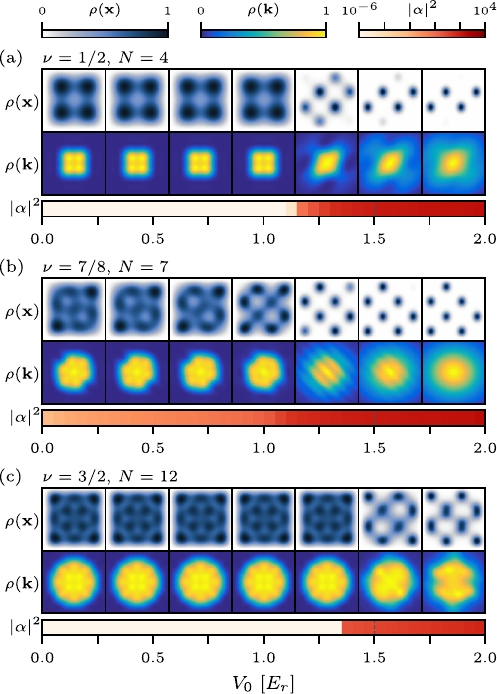}
  \caption{ 
  % {\color{magenta}[Update caption]}
  Three different scenarios observed for Pauli crystal superradiance for different fillings $\nu$. The real ($\rho(\vb{x})$) and momentum space densities ($\rho(\vb{k})$) of Pauli crystals at $\nu=1/2,7/8,3/2$ are shown together with the values of $|\alpha|^2$ in log scale (colorbars) at different values of $V_0$. The densities are shown in $x, y \in [0, 2\lambda]$ and $k_x, k_y \in [-4\pi/\lambda, 4\pi/\lambda]$, and normalized to the corresponding maximum value. The blue dashed lines in the $|\alpha|^2$ colorbar at $V_0 =1.5\,E_r$ indicate the bosonic threshold. \textbf{(a)} $\nu=1/2,N=4$ {\it closed-shell}: Superradiance is achieved beyond critical pump strength. \textbf{(b)}  $\nu=7/8,N=7$ {\it open-shell subject to cavity}: Soft transition is achieved at vanishing pump strength, while the system localizes only at large pump strengths. \textbf{(c)} $\nu=3/2,N=12$: {\it open-shell not subject to cavity}: Soft transition is not achieved, and the system enters superradiance only at strong pump strengths.
  }

  \label{fig:fig3}
\end{figure}

{\it  Closed-shell Pauli crystals}: 
We first consider the specific case of $\nu=\frac12$, $N=4$, where superradiance is expected to be induced by the recoil process $\ket{2,2} \to (\ket{1,3} \pm \ket{3,1})/\sqrt{2}$ at a finite critical point.
Our numerical simulations indeed confirm the absence of a soft transition. 
The real and momentum space densities at different $V_0$ are summarized in Fig.~\ref{fig:fig3}(a). As $V_0$ increases from 0 to $\sim1.15E_r$ just below the superradiant threshold, both the Pauli crystal in position and momentum space remain unaffected. The onset of superradiance takes place already at $V_0\sim1.15E_r$, earlier than the bosonic threshold (dashed blue line in Fig.~\ref{fig:fig2}). 
Consequently, the atoms effectively deform in  momentum space along one of the two $\vb{q}_\pm$ vectors; the extent of this deformation is now comparable to its magnitude $\sim q$ and the cavity becomes populated by a coherent state. 
As $V_0$ increases,  the atoms  scatter to  even higher eigenmodes  and contribute to  superradiance.
Finally, at large $V_0\sim1.8E_r$ each fermion localizes in an individual minimum of the scattering lattice with a concomitant delocalization in $\rho(\kk)$.  These conclusions qualitatively hold for all closed-shell configurations irrespective of the specfic fermion density. 
Our simulations show that this threshold has a V-shaped dependence on the filling $\nu$ with the required critical coupling exhibiting a minimum at $\nu=1$. This feature is compatible with Refs.~\cite{piazza2014Umklapp, chen2014superradiance}.

{\it Open-shell Pauli crystals }: In our implementation with $\beta=4$ and within our range of particle numbers, open shell configurations arise at the following filling fractions: $\nu= \frac28,\frac58,\frac78,\frac98,\frac{12}{8},\frac{14}{8}$. 
Among these configurations, the cavity field mediates soft superradiant transitions when the most energetic fermion occupies eigenmodes $|n_a,n_b\rangle$ (or $|n_b,n_a\rangle$) with $n_a+n_b$ being odd. This includes: $\nu=\frac14$, $N=2$ where $(n_a=1, n_b=2)$, $\nu=\frac78$, $N=7$ with $(n_a=3, n_b=2)$, $\nu=\frac98$ with  $(n_a=4, n_b=1)$ and $\nu=\frac74$, $N=14$ with  $(n_a=3, n_b=4)$. 
In Fig.~\ref{fig:fig1}(e), we show the strength of $W$ connecting the different elements on the momentum lattice. We see that its magnitude is maximal for states in the vicinity of $\nu=1$ namely fractions $\nu=\frac78, \frac98$ and decreases  for states with higher quantum numbers.
Focusing on numerical simulations for $\nu=\frac78$, $N=7$, we obtain a small yet non-zero value of \( |\alpha|^2 \sim \order{10^{-3}} \) at zero threshold [Fig. \ref{fig:fig3}(b)].  As $V_0$ increases, the system  smoothly crosses over into the superradiant phase where $\alpha$ grows and the atomic distributions   deform while largely preserving the  Pauli crystalline structure. This is in contrast to the usual second-order  superradiant phase transition which is accompanied by diverging fluctuations stemming from spontaneous symmetry breaking.  Consequently, cavity fluctuations do not  diverge as $V_0$ approaches zero for the soft transitions. Note that beyond  $V_0\sim 1.25 E_r$, the Pauli crystal vastly deforms as scattering to higher  eigenmodes dominates  and  the distinction between open- and closed-shell configurations become irrelevant.  

For all the other open-shell configurations with even $n_a \pm n_b$, the cavity does not mediate soft superradiant transitions due to destructive interference $W_{n_a,n_b}=0$ and thus lack of Fermi sea squeezing. In this case, the superradiance is again driven by the usual Dicke-type processes at finite critical couplings.
Consider $\nu=\frac32$, $N=12$, shown in Fig.~\ref{fig:fig3}(c) as an example. The open shell Pauli crystal remains in the normal phase at small pump rates, because the soft transition is forbidden by destructive interference between  $|2,4\rangle$ and $|4,2\rangle$. The onset of superradiance is mediated  by the scattering from $(|1,3\rangle\pm|3,1\rangle)/\sqrt{2}$ to $(|2,4\rangle\pm|4,2\rangle)/\sqrt{2}$, accompanied by macroscopic values for $|\alpha|^2$. As a result, the Pauli crystal deforms into a dumbbell-shaped momentum density, as seen at $V_0\sim 1.5E_r$.

We emphasize that these soft transitions are intrinsically precluded in one dimension, where  the fermions only  occupy closed-shell and hence non-degenerate Pauli crystal configurations.
Moreover, these soft transitions are unrelated to Umklapp superradiance \cite{piazza2014Umklapp, Keeling2014Fermionic, chen2014superradiance}, and do not require any fine tuning of the cavity wavelength.

Our theoretical predictions  can be tested  in experimental setups, as all relevant ingredients are available state-of-the-art technologies, even if not yet in the required combination. This includes essentially the Pauli crystal via strong  confinement~\cite{holten2021observation}, the square box potential via programmable spatial light modulators and digital micromirror devices~\cite{navon2021quantum}, and the cavity-fermion coupling~\cite{zhang2021observation,helson2023densitywave}. Moreover, our predictions  are qualitatively robust against additional perturbations pertaining to realistic experimental realizations, e.g., a transverse pump standing wave optical lattice potential $V_{p}=V_0 \sin^2(\vb{k}_p\vb{x})$. 
As shown in SM~\cite{supmat}, this potential does not have any qualitative impact on the soft transitions in the regime $\nu <1$, but eliminates all soft transitions for $\nu >1$,  recovering the usual finite threshold behavior for  superradiance. This can be inferred from the fact that when $\nu >1$, this optical lattice potential induces an extra energy cost  as the fermion distributions get denser, eliminating the degeneracy.

In summary, this Letter highlights novel emergent quantum many-body phenomena in  a cavity-fermion system stemming from the interplay between Fermi statistics and confinement at the single particle level. Building on the elementary physics of a quantum particle in a box,   we unveil a pathway to zero-threshold ordered phases beyond the Landau paradigm. 
Extensions to systems with internal degrees of freedom, tunable interactions, or spin-orbit coupling offer rich opportunities to uncover unconventional phases and collective excitations. Of particular interest is the potential to engineer massive Nambu–Goldstone modes~\cite{watanabe2013massive,ohashi2017conformal} by extending the symmetry structure to continuous groups such as $\mathrm{U}(1)$, e.g., via crossed-cavity architectures~\cite{leonard2017monitoring}. These findings lay the groundwork for future experiments in quantum gas–cavity platforms, where the effects of confinement, quantum statistics, and light-matter coupling can be precisely controlled.

\begin{acknowledgments}
    D.O.-G. acknowledges funding from the ETH Research Grant. R.L. acknowledges funding from the Innovation Program for Quantum Science and Technology under grant number 2024ZD0301800. T.D. acknowledges funding from the Swiss State Secretariat for Education, Research and Innovation (SERI) under grant number MB22.00090
    and from the Swiss National Science Foundation (SNSF): project numbers 217124, 221538, 223274.
\end{acknowledgments}

\bibliography{apssamp}% Produces the bibliography via BibTeX.

\widetext

\newpage

\begin{center}
  \textbf{\large Supplemental Material: \\ Pauli crystal superradiance}
  \vspace{20pt}
\end{center}

%%%%%%%%%% Prefix a "S" to all equations, figures, tables and reset the counter %%%%%%%%%%
\setcounter{equation}{0}
\setcounter{figure}{0}
\setcounter{table}{0}
\setcounter{section}{0}
\makeatletter
\renewcommand{\theequation}{S\arabic{equation}}
\renewcommand{\thefigure}{S\arabic{figure}}
\renewcommand{\bibnumfmt}[1]{[S#1]}
% \renewcommand{\citenumfont}[1]{S#1}
%%%%%%%%%% Prefix a "S" to all equations, figures, tables and reset the counter %%%%%%%%%%

\twocolumngrid

\section{Pauli crystals of a box} \label{sec:S_PCbox}

Pauli crystals are unique geometric structures formed solely by the interplay of kinetic energy, a trapping potential, and the Pauli exclusion principle. As more fermions are added to a trap, Fermi statistics results in the emergence of distinct geometric structures in the ground state. These  Pauli crystals exhibit progressively smaller inter-particle spacing with increasing particle number, leading to a more spread-out momentum distribution. Furthermore, symmetries of the trap lead to  degeneracies and hence a continuum of equivalent Pauli crystal structures. This aspect is highly relevant for the soft transitions discussed in the main text. 

We now discuss  features of Pauli crystals in both position and momentum space for a square box of length $L$. We start by writing down the eigenstates of the trap.  As the Schrödinger equation is separable in the  coordinate projections $x$ and $y$, it suffices to obtain    position and momentum space densities
in 1D.  For a box of length $L$,  the eigenstates in position and momentum space are given by ($n=1,2,3,\dots$)
\begin{align}
\psi_{n}(x) =& \sqrt{\frac{2}{L}} \sin(\frac{n\pi}{L} x), \\
\psi_n(k) =& \frac{1}{2i}\sqrt{\frac{L}{\pi}} \Bigg[ (-1)^n\text{sinc}\left(\frac{kL-n\pi}{2}\right) \nonumber  \\ 
& - \text{sinc}\left(\frac{kL+n\pi}{2}\right) \Bigg]e^{ikL/2}, \label{eq:S_1Dk}
\end{align}
where $\text{sinc}(x) = \sin(x)/x$ is the cardinal sine function, which satisfies $\lim_{L\rightarrow\infty}L \text{ sinc}(kL)= \delta(k)$. The 2D eigenfunctions are  $\phi_{\underline{n}}(\vb{x}) =\psi_{n_x}(x)\psi_{n_y}(y)$ with the label ${\underline{n}}\equiv (n_x,n_y)$ ordered by increasing eigenenergy 
%{\color{magenta}[Is it really necessary to introduce the 2D $n$, instead of keeping using $n_x,n_y$? It can induce confusion.]}
\begin{equation}
E_{\underline{n}} = \frac{(n_x^2+n_y^2)\pi^2}{2mL^2} \label{eq:S_En}.
\end{equation}
For example,  $\underline{n}=\underline{1}$ corresponds to $(1,1)$, $\underline{n}=\underline{2}$ to $(2,1)$, $\underline{n}=\underline{4}$ to $(2,2)$, and so on as seen in Fig. \ref{fig:S_spectrum_fill}. Note that for some particle numbers we have a single (multiple) ground state(s), where the most energetic fermion lies in a closed (open) energy shell or level (see Fig. \ref{fig:S_spectrum_fill}a).  For example, for $N=2$, the ground state populates with its first fermion the state (1,1) but with the second it can populate either the state (1,2), the (2,1) or a superposition of both of them.

\begin{figure}[htb]
  \includegraphics[width=1\columnwidth]{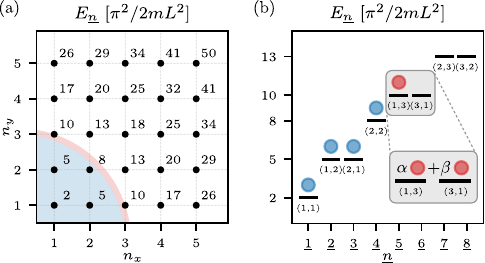}
  \caption{ \textbf{(a)} Eigenenergies for the single-particle states in a 2D square box $\phi_{\underline{n}}(\vb{x})$. The blue circle encloses the occupied states of the ground state of $N=4$ fermions in a box; the red circle encloses the same states plus a superposition of $(1,3)$ and $(3,1)$. \textbf{(b)} Energy ladder diagram of fermions in a 2D box. The blue and red balls are associated to the circles and populations in (a). 
  }
  \label{fig:S_spectrum_fill}
\end{figure} 

\begin{figure*}
  \includegraphics[width=\linewidth, trim={0 2mm 0 0},clip]{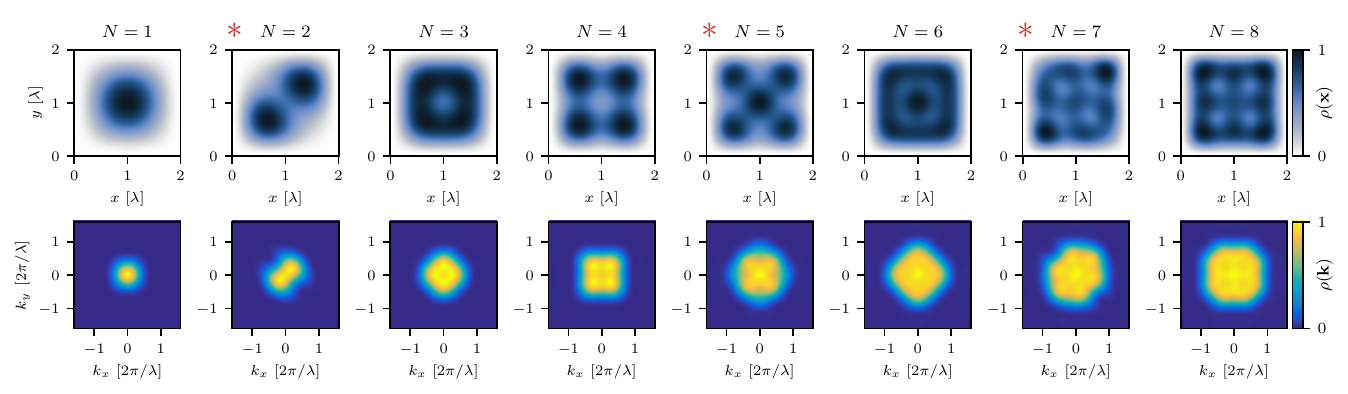}
  \includegraphics[width=\linewidth]{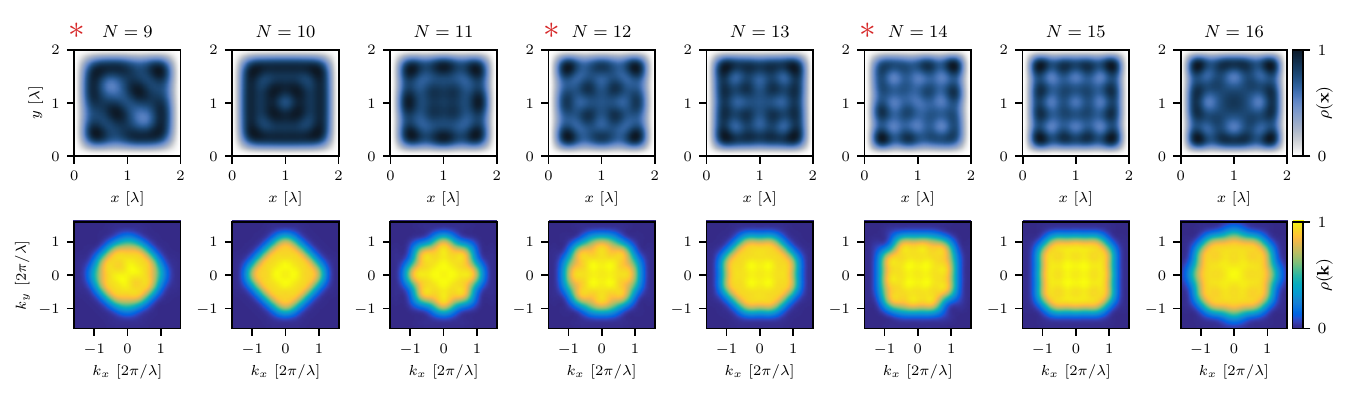}
    % \vspace{-20pt}
  \caption{ Analytically calculated Pauli crystals of $N=1$ to $N=16$ fermions in a 2D box  with $x, y \in [0, 2\lambda]$ in position (top) and momentum (bottom) space. Particle numbers marked with a red $\ast$ are open-shell Pauli crystals with the most energetic fermion occupying occupying a superposition of configurations, where one possibility is shown here. In particular, the symmetric superposition of eigenmodes $(\ket{n_a,n_b} + \ket{n_b, n_a})/\sqrt{2}$ takes the following values: We have for $N=2$: $n_a = 1$, $n_b=2$, for $N=3$: $n_a = 1$, $n_b=3$, for $N=5$: $n_a = 1$, $n_b=3$, for $N=7$: $n_a = 2$, $n_b=3$, for $N=9$: $n_a = 1$, $n_b=4$, for $N=12$: $n_a = 2$, $n_b=4$ and for $N=14$: $n_a = 3$, $n_b=4$. For $n_a+n_b$ odd, both $\rho(\vb{x})$ and $\rho(\vb{k})$  are elongated/squeezed along the positive diagonal, while for $n_a+n_b$ even, they are equally deformed along both diagonals. All densities have been normalized to their maximum value. }
  \label{fig:SboxPC}
  \end{figure*}

The full $N$-particle wavefunction can be written as a  Slater determinant of the states $\phi_{\underline{n}}$.  Moreover, it can be shown that the one-body density matrix $\rho(\vb{x}, \vb{x}')$ and its diagonal, the one-body density $\rho(\vb{x})$, for such a groundstate simplify to 
\begin{align}
\rho(\vb{x}, \vb{x}')  =& \frac{1}{N}\sum_{n= 1 }^{N} \phi^*_{\underline{n} }(\vb{x})\phi_{\underline{n} }(\vb{x}') , \label{eq: 1corr}\\
\rho(\vb{x})  =& \frac{1}{N}\sum_{n = 1 }^{N} |\phi_{\underline{n}}(\vb{x})|^2.\label{eq: 1rho}
\end{align}
The momentum space equivalent has the same form wherein  $\phi_{\underline{n} }(\vb{x}) \rightarrow \phi_{\underline{n} }(\kk)$.  The Pauli crystal densities   for a square box of size $L=2\lambda$ and varying particle number $N$  are shown  in Fig. \ref{fig:SboxPC}. However, the open-shell Pauli crystals marked with a red $\ast$  have  multiple allowed  configurations given by $\rho(\vb{x})$ and $\rho(\kk)$ for a given particle number. In the next section, we show this has an effect on the soft-transition superradiance presented in the main text.

\section{Degeneracy of Pauli crystals and superradiance} \label{sec:S_deg}

\begin{figure*}
  \includegraphics[width=1\linewidth]{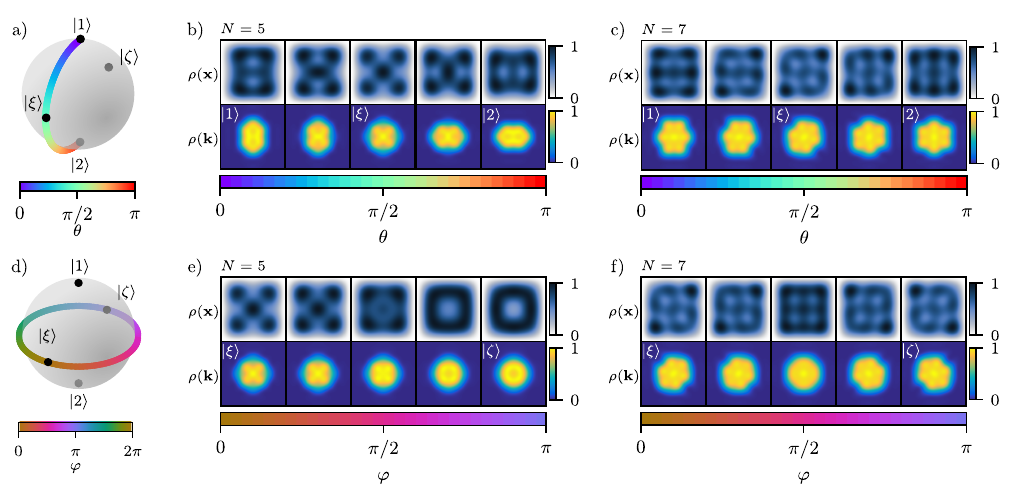}
  \caption{ Multiple Pauli crystal configurations due to a two-fold degenerate energy level (Analytically calculated). \textbf{(a)} Given two degenerate Pauli crystals $\ket{1}$ and $\ket{2}$ there exists a continuum of Pauli crystals that can be formed on the Bloch sphere. Antipodal states $\ket{\xi}$ and $\ket{\zeta}$ are orthogonal. Consider a trajectory on the Bloch sphere for the polar angle $\theta\in [0,\pi]$. \textbf{(b,c)} Continuous set of Pauli crystals for $N=5, 7$ fermions in the Bloch sphere from (a) as a function of the polar angle $\theta$. In both particle numbers, the crystals associated to $\ket{1}$ and $\ket{2}$ are the same up to $x\leftrightarrow y$. \textbf{(d)} Consider a trajectory in the Bloch sphere for the azimuthal angle $\varphi \in [0,2\pi)$. \textbf{(e,f)} Continuous set of Pauli crystals for $N=5,7$ fermions in the Bloch sphere from (a) in the equator as a function of the azimuthal angle $\varphi$. In all cases the position and momentum one-body densities ($\rho(\vb{x})$ and $\rho(\vb{k})$) were normalized to have their maximum value being 1. $\rho(\vb{x})$ plotted in range $x,y\in[0,L]$ and $\rho(\vb{k})$ plotted in range $k_x,k_y\in[-6\pi/L,6\pi/L]$.
  }
  \label{fig:S_bloch}
\end{figure*} 

Here we  focus on the open shell Pauli crystals and  their implications for  the zero-threshold  soft-transition superradiance from the main text. Since we deal with degenerate energy levels,  we use perturbation theory to see which crystal is `selected' by the cavity. 

We consider a system in a trap with $N$ particles, where the first $N-1$ particles completely fill the lower energy levels and the last particle lies in an empty two-fold degenerate energy level. An example of this case is the one of a square box like the one from the main text or the previous section. However, the treatment here is more general and could be further generalized to larger degeneracies. We denote by $\ket{1}$ and $\ket{2}$ the many-body  degenerate states of the $N$ particles, $\ket{1} = \ccd[d1]\prod_{\ell=1}^{N-1}\ccd[\ell]\ket{\text{vac}}$ and   $\ket{2} = \ccd[d2]\prod_{\ell=1}^{N-1}\ccd[\ell]\ket{\text{vac}}$  where  $\ccd[i]$ is a fermion creation operator in the $i$-th single particle  state and [$d1$] and  [$d2$] index the  degenerate  single particle states in the last shell. { The corresponding Pauli crystal densities can be written in the single particle basis $\phi_{\underline{\ell}} (\vb{x})$ as } 
\begin{equation}
  \rho_{1,2}(\vb{x}) = \frac{1}{N} \left[ |\phi_{\underline{d1},\underline{d2}}(\vb{x})|^2 + \sum_{\ell=1}^{N-1}|\phi_{\underline{\ell}}(\vb{x})|^2\right].
\end{equation}
These are two possible Pauli crystal but we can construct a continuum of them by superpositions of the two last levels. 

The interaction Hamiltonian takes the form
\begin{equation}
  \hat{V} = \int \dd\vb{x} \FOdx \sqrt{U_0V_0} (\ca+\cad)\sin(\vb{k}_p\vb{x})\sin(\vb{k}_c\vb{x})\FOx. \label{eq:S_Vint}
\end{equation}
Within a degenerate subspace of $\Ha[0]$ such as that spanned by $\{\ket{1}, \ket{2}\}$, this Hamiltonian takes the form of the matrix
\begin{equation}
  \hat{V} = \mqty(\mel*{1}{\hat{V}}{1} & \mel*{1}{\hat{V}}{2} \\
  \mel*{2}{\hat{V}}{1} &\mel*{2}{\hat{V}}{2}), \label{eq:S_Vmat}
\end{equation}
introducing an effective coupling between the two degenerate basis states, and thus lifting the degeneracy.
Upon diagonalization in this subspace, we obtain the following orthonormal superposition states as new basis states,
\begin{align}
  \ket{\xi(\theta,\varphi)} =&  \cos\frac{\theta}{2}\ket{1} +e^{i\varphi}\sin\frac{\theta}{2} \ket{2},\\
  \ket{\zeta(\theta,\varphi)} =&  \sin\frac{\theta}{2}\ket{1}-e^{i\varphi}\cos\frac{\theta}{2} \ket{2},
\end{align}
which fulfill $\braket{\xi}{\zeta}=0$ and can be mapped to antipodal points in a Bloch sphere [see Fig. \ref{fig:S_bloch}].  Here, $\ket{\xi}$ and $\ket{\zeta}$ are also degenerate crystals in the absence of cavity coupling. In this basis, the matrix from Eq. (\ref{eq:S_Vmat}) can be rewritten as 
\begin{equation}
  \mqty(\mel*{\xi}{\hat{V}}{\xi} & \mel*{\xi}{\hat{V}}{\zeta} \\
  \mel*{\zeta}{\hat{V}}{\xi} &\mel*{\zeta}{\hat{V}}{\zeta}), \label{eq:S_Vmat2}
\end{equation}
where we have the freedom  to choose $(\theta, \varphi)$ such that the  off-diagonal terms are zero.  For our setup of  a square box, symmetry  arguments indicate that  $\ket{1} \to \ket{2}$  when the spatial coordinates  $x\leftrightarrow y$.  Consequently, $\ev*{\hat{V}}{1}=\ev*{\hat{V}}{2}$, which in conjunction with the  diagonalization requirement $\mel*{\xi}{\hat{V}}{\zeta} \overset{!}{=}0$ leads to $\theta=\pi/2$. Now, if we want to maximize the energy splitting 
\begin{align}
  \mel*{\xi}{\hat{V}}{\xi} & = \ev*{\hat{V}}{1} + \Re{e^{i\varphi}\mel*{1}{\hat{V}}{2}},  \label{eq:S_xi}\\
  \mel*{\zeta}{\hat{V}}{\zeta} & = \ev*{\hat{V}}{2} - \Re{e^{i\varphi}\mel*{1}{\hat{V}}{2}}, \label{eq:S_zeta}
\end{align}
we need to choose $\varphi$ accordingly.  These expressions are further simplified  by noting that both the potential and the wavefunctions are oscillatory functions with a wavelength commensurate with $L$.  Defining $\Delta \varphi =\arg{\mel*{1}{\hat{V}}{2}}$, we find  that $\varphi = -\Delta \varphi$ and $\varphi=\pi-\Delta \varphi$ give the maximal energy splitting  {\it if and when } $\mel*{1}{\hat{V}}{2}\neq 0$.

The fact that $\varphi$  has two values   reflects the  two possible $\mathbb{Z}_2$ symmetry-broken states corresponding to $\ev{\ca}= \pm \alpha$.  If we consider the full quantum problem with the cavity in its ground state $\ket{0}_{\text{c}}$,  the original two-fold degenerate states [Pauli crystals] $\ket{1}_{\text{a}}\otimes\ket{0}_{\text{c}}$ and $\ket{2}_{\text{a}}\otimes\ket{0}_{\text{c}}$ split into four states in the soft-transition case: two ground states $\ket{\zeta}_{\text{a}}\otimes\ket{\alpha}_{\text{c}}$ and $\ket{\xi}_{\text{a}}\otimes\ket{-\alpha}_{\text{c}}$ and two excited states $\ket{\zeta}_{\text{a}}\otimes\ket{-\alpha}_{\text{c}}$ and $\ket{\xi}_{\text{a}}\otimes\ket{\alpha}_{\text{c}}$  where $\ket{\alpha}$ is a coherent state of the cavity. This  splitting  has important consequences for our system. Typically, a symmetry broken phase transition is accompanied by the closing of a gap (or dissipative gap in the open system case) \cite{minganti2018spectral}. The reason behind this lies in the dimensionality of  the asymptotic subspace of the ground states (steady states of the  dissipative system):  before the symmetry breaking  transition this space is one-dimensional [spanned by a single ground state] and after a $\mathbb{Z}_n$ symmetry is broken it should become $n$-dimensional [spanned by $n$ ground states] \cite{minganti2018spectral, lieu2020symmetry}. However, in the soft-transition cases  we have, the symmetry is immediately broken at zero threshold and  the number of states spanning the asymptotic subspace remains two across this transition.
This reflects the absence of any gap closure at the transition  and hence the absence of cavity fluctuations in the soft-transition case  as opposed to the usual  superradiant transitions at a finite non-zero threshold at  other filling fractions.

\section{Calculation of W and superradiance in a box}
As we saw in the previous section, the condition of $\mel*{1}{\hat{V}}{2}\neq 0$ leads to a degeneracy lifting of Pauli crystals in the presence of a minimal interaction Hamiltonian of the form of the lattice associated to superradiance [See Eq. (\ref{eq:S_Vint})]. In particular, we can directly calculate all many-body matrix elements in terms of single-particle quantities. To do this we expand the field-operator as $\FOx = \sum_{n=1}^{\infty} \phi_{\underline{n}}(\vb{x}) \cc[n]$ and use Wick's theorem given our Slater determinant state to obtain sums of individual integrals, which we express as 
\begin{align}
  \ev*{\hat{V}}{1}= & \ev*{\hat{V}^{(1)}}{\underline{d1}} + \sum_{n=1}^{N-1} \ev*{\hat{V}^{(1)}}{\underline{n}} \label{eq:S_1V1}\\
  \ev*{\hat{V}}{2}= &  \ev*{\hat{V}^{(1)}}{\underline{d2}} +  \sum_{n=1}^{N-1} \ev*{\hat{V}^{(1)}}{\underline{n}}\\
  \mel*{1}{\hat{V}}{2} = &  \mel*{\underline{d1}}{\hat{V}^{(1)}}{\underline{d2}}, \label{eq:S_1V2}
\end{align}
where $\hat{V}^{(1)} =\sqrt{U_0 V_0} (\ca+\cad) \sin(\vb{k}_p \hat{\vb{x}})\sin(\vb{k}_c\hat{\vb{x}})$ is the single-particle interacting Hamiltonian in first quantization and $\ket{\underline{n}}$ is the single-particle $n$-th eigenstate of the square box. Notice however that $\forall n$  we must have $\ev*{\hat{V}^{(1)}}{\underline{n}}=0$, since otherwise even a single fermion in a 2D box has a non-zero expectation value of the order parameter for superradiance and thus trivially pre-breaks the $\mathbb{Z}_2$ symmetry.

Nonetheless, it is clear that our perturbation depends on matrix elements of the form
\begin{equation}
  \mel*{\underline{n}}{ \sin(\vb{k}_p \hat{\vb{x}})\sin(\vb{k}_c\hat{\vb{x}})}{\underline{m}} = W_{n_x, m_x}W_{n_y, m_y},
\end{equation}
where we have used the 90$^\circ$ transverse pump - cavity configuration to split the matrix element into 1D single-particle matrix elements
\begin{align}
  W_{n, m} =&  \mel*{n}{\sin(2\pi \hat{x}/\lambda)}{m}\\
  =& \frac{2}{L} \int_0^L \dd x \sin(\frac{n \pi x}{L}) \sin(\frac{2\pi x}{\lambda}) \sin(\frac{m \pi x}{L}), 
\end{align}
with $|\vb{k}_{p(c)}| = 2\pi/\lambda$. 

\begin{table}[h]
    \centering
    \renewcommand{\arraystretch}{1.3}  % Stretch rows
    \setlength{\tabcolsep}{0pt}       % Column spacing
    \begin{tabular}{cc|ccccccccc}
        % \toprule
        & & \multicolumn{9}{c}{$m$}  \\
        % \cmidrule(r){3-11}
        &$W_{n,m}$ & 1 & 2 & 3 & 4 & 5 & 6 & 7 & 8 & 9  \\ 
        \hline
        \multirow{9}{*}{$n$} \hspace{-2mm}
        & 1  & $0$ & $-\frac{64}{105\pi}$ & $0$ & $\frac{128}{63\pi}$ & $0$ & $-\frac{64}{99\pi}$ & $0$ & $-\frac{256}{2145\pi}$ & $0$ \\
        & 2  & $-\frac{64}{105\pi}$ & $0$ & $\frac{64}{45\pi}$ & $0$ & $\frac{320}{231\pi}$ & $0$ & $-\frac{448}{585\pi}$ & $0$ & $-\frac{64}{385\pi}$ \\
        & 3  & $0$ & $\frac{64}{45\pi}$ & $0$ & $\frac{128}{165\pi}$ & $0$ & $\frac{576}{455\pi}$ & $0$ & $-\frac{256}{315\pi}$ & $0$ \\
        & 4  & $\frac{128}{63\pi}$ & $0$ & $\frac{128}{165\pi}$ & $0$ & $\frac{128}{195\pi}$ & $0$ & $\frac{128}{105\pi}$ & $0$ & $-\frac{128}{153\pi}$ \\
        & 5  & $0$ & $\frac{320}{231\pi}$ & $0$ & $\frac{128}{195\pi}$ & $0$ & $\frac{64}{105\pi}$ & $0$ & $\frac{1280}{1071\pi}$ & $0$ \\
        & 6  & $-\frac{64}{99\pi}$ & $0$ & $\frac{576}{455\pi}$ & $0$ & $\frac{64}{105\pi}$ & $0$ & $\frac{448}{765\pi}$ & $0$ & $\frac{1728}{1463\pi}$ \\
        & 7  & $0$ & $-\frac{448}{585\pi}$ & $0$ & $\frac{128}{105\pi}$ & $0$ & $\frac{448}{765\pi}$ & $0$ & $\frac{1792}{3135\pi}$ & $0$ \\
        & 8  & $-\frac{256}{2145\pi}$ & $0$ & $-\frac{256}{315\pi}$ & $0$ & $\frac{1280}{1071\pi}$ & $0$ & $\frac{1792}{3135\pi}$ & $0$ & $\frac{256}{455\pi}$ \\
        & 9  & $0$ & $-\frac{64}{385\pi}$ & $0$ & $-\frac{128}{153\pi}$ & $0$ & $\frac{1728}{1463\pi}$ & $0$ & $\frac{256}{455\pi}$ & $0$ \\
        % \bottomrule
    \end{tabular}
    \caption{Matrix with values of $W_{n,m}(\beta)$ for $\beta=4$ ($L=2\lambda$).}
    \label{tab:Wnmbeta4}
\end{table}

\begin{table}[h]
    \centering
    \renewcommand{\arraystretch}{1.3}  % 1.5x default row height
    \setlength{\tabcolsep}{5.5pt}
    \begin{tabular}{cc|ccccccccc}
        % \toprule
        & & \multicolumn{9}{c}{$m$}  \\
        % \cmidrule(r){3-11}
        &$Y_{n,m}$ & 1 & 2 & 3 & 4 & 5 & 6 & 7 & 8 & 9  \\ 
        \hline
        \multirow{9}{*}{$n$} \hspace{-2mm}
        & 1  & 0 & 0 & $-\frac{1}{2}$ & 0 & $\frac{1}{2}$ & 0 & 0 & 0 & 0 \\
        & 2 & 0 & $-\frac{1}{2}$ & 0 & 0 & 0 & $\frac{1}{2}$ & 0 & 0 & 0 \\
        & 3 & $-\frac{1}{2}$ & 0 & 0 & 0 & 0 & 0 & $\frac{1}{2}$ & 0 & 0 \\
        & 4 & 0 & 0 & 0 & 0 & 0 & 0 & 0 & $\frac{1}{2}$ & 0 \\
        & 5 & $\frac{1}{2}$ & 0 & 0 & 0 & 0 & 0 & 0 & 0 & $\frac{1}{2}$ \\
        & 6 & 0 & $\frac{1}{2}$ & 0 & 0 & 0 & 0 & 0 & 0 & 0 \\
        & 7 & 0 & 0 & $\frac{1}{2}$ & 0 & 0 & 0 & 0 & 0 & 0 \\
        & 8 & 0 & 0 & 0 & $\frac{1}{2}$ & 0 & 0 & 0 & 0 & 0 \\
        & 9 & 0 & 0 & 0 & 0 & $\frac{1}{2}$ & 0 & 0 & 0 & 0 \\
        % \bottomrule
    \end{tabular}
    \caption{Matrix with values of $Y_{n,m}(\beta)$ for $\beta=4$ ($L=2\lambda$).}
    \label{tab:Ynmbeta4}
\end{table}

This matrix element clearly depends on the relation of $L$ to $\lambda$. We thus define a measure of our box size $\beta = 2L/\lambda \in \mathbb{Z}$, such that $W_{n, m}(\beta)$ has encoded within it the dependency of the system's size in a single parameter. This quantity is easily calculated by introducing two other matrix elements. Let 
\begin{equation}
  \mel*{n}{e^{ i \pi \beta \hat{x}/L}}{m} = Y_{n, m}(\beta)+ i W_{n, m}(\beta), \label{eq:S_ei}
\end{equation}
where we defined 
\begin{equation}
  Y_{n, m}(\beta) =  \mel*{n}{\cos(2\pi \hat{x}/\lambda)}{m}
\end{equation}
as the even part (as a function of $\beta$) of Eq. (\ref{eq:S_ei}), while $W_{n, m}(\beta)$ is its odd part. Tables \ref{tab:Wnmbeta4} and \ref{tab:Ynmbeta4} contain some explicit values for $W_{n,m}(\beta=4)$ and $Y_{n,m}(\beta=4)$. Notice that the maximum magnitudes of $W_{n,m}$ are around the lines $n+m=\beta$ and $|n-m|=\beta$.

We can directly calculate Eq. (\ref{eq:S_ei}) by using $u = x/L$ and $\sin( \pi n  u) = \sum_{s=\pm} s \exp(i\pi (s n) u )/2i$:
\begin{align}
  \mel*{n}{e^{ i \frac{\pi \beta \hat{x}}{L}}}{m} = &  \sum_{s, s' = \pm } -\frac{s s'}{2} \int_0^1 \dd u e^{i \pi (s n+s' m+\beta) u } \\
  = &  \sum_{s, s' = \pm } \frac{s s'}{2}  \frac{1-e^{i \pi (s n+s' m+\beta) }}{i \pi (s n+s' m+\beta)} \nonumber \\
  & - \sum_{s, s' = \pm } \frac{s s'}{2} \delta_{s n+s' m+\beta, 0},
\end{align}
where we have introduced a Kronecker delta for the cases when $s n+s' m+\beta=0$. Moreover, we can simplify the expression by introducing $\sigma \equiv s s'$, summing over $s$ and noticing $e^{i s (n+\sigma m )}= (-1)^{n+\sigma m }= (-1)^{n+ m }$ for $s \in\{ \pm\}$ and $n,m\in \mathbb{Z}$
\begin{align}
  \mel*{n}{e^{ i \frac{\pi \beta \hat{x}}{L}}}{m} = &  \sum_{\sigma = \pm } \frac{\sigma \beta}{i\pi}  \frac{(1- (-1)^{n+ m +\beta } )}{ \beta^2 - (n+\sigma m)^2} \nonumber \\
  & - \sum_{\sigma  = \pm } \frac{\sigma }{2} \delta_{|n+\sigma m|,|\beta| }\\
  = &  i \frac{\beta}{\pi} \frac{8 n m \text{ }  \Pi_{n\pm m , \beta}}{(\beta^2 -n^2-m^2)^2 - 4n^2 m^2} \nonumber \\
  & - \frac{1}{2} \delta_{|n+ m|,|\beta| } + \frac{1}{2} \delta_{|n- m|,|\beta| },
\end{align}
where the first term is $ i W_{n, m}(\beta)$ and the rest is $Y_{n, m}(\beta)$ and we have introduced 
\begin{equation}
  \Pi_{a,b} = (a+b) \ \mathrm{ mod } \ 2,
\end{equation}
a `parity selector', which is only non-zero when $n\pm m$ has a parity different to that of $\beta$. 

Notice this has an important implication: $\beta$ has to be an even number. If this were not the case, $\ev*{\hat{V}^{(1)}}{\underline{n}} \neq 0$ already at $n=1$, therefore trivially pre-breaking the $\mathbb{Z}_2$ symmetry for any particle number. We thus restrict ourselves to the non-trivially broken $\mathbb{Z}_2$, where we have
\begin{align}
  W_{n, m}(\beta) = &\frac{4 [1-(-1)^{n\pm m}] n m }{(\beta^2 -n^2-m^2)^2 - 4n^2 m^2} \label{eq:S_Wnm}\\
  Y_{n, m}(\beta) = &  - \frac{1}{2} \delta_{|n+ m|,|\beta| } + \frac{1}{2} \delta_{|n- m|,|\beta| }.  \label{eq:S_Ynm}
\end{align}

Using these expressions we can obtain specific expressions for the matrix elements of $\hat{V}$ in the degenerate subspace of $\Ha[0]$ [cf. Eqs. (\ref{eq:S_1V1}-\ref{eq:S_1V2})]. We note that in our particular case of an open energy shell in a square box, we have $\ket{\underline{d1}} = \ket{n_x, n_y} = \ket{a,b}$  and $\ket{\underline{d2}} = \ket{b,a}$. Hence we obtain $\mel*{1}{\hat{V}}{2} =\mel*{2}{\hat{V}}{1}  = \sqrt{U_0 V_0} (\ca+\cad) W_{a,b}^2(\beta)$ and $\mel*{1}{\hat{V}}{1} = \mel*{2}{\hat{V}}{2}=0$. Clearly, by Eqs. (\ref{eq:S_xi}-\ref{eq:S_zeta})
\begin{equation}
  \ev{ \hat{V} }{\pm}_{\text{a}}  = \pm\sqrt{U_0 V_0} (\ca+\cad)  W_{a,b}^2(\beta), \label{eq:S_Va}
\end{equation}
leading to a maximal splitting with states $\ket{\xi} = \ket{+}$ and $\ket{\zeta} = \ket{-}$ where
\begin{align}
  \ket{\pm} = & \frac{1}{\sqrt{2}}\left( \ket{1} \pm \ket{2} \right),
\end{align}
which makes the most energetic fermion populate the open-shell in a superposition state $\propto \ket{a,b}\pm \ket{b,a}$. This leads to the superradiant states at zero $V_0$ threshold  $\ket{+}_{\text{a}}\otimes\ket{-\alpha}_{\text{c}}$ and  $\ket{-}_{\text{a}}\otimes\ket{\alpha}_{\text{c}}$, i.e. the (anti-)symmetric superposition of Pauli crystals will correspond to a $\mathbb{Z}_2$ broken superradiant state with (positive) negative phase of light. 

Nonetheless, this is a finite size effect induced by the confining potential. It is clear from  Eq.(\ref{eq:S_Wnm}) and $\beta= 2L/\lambda$ that $\lim_{L\rightarrow \infty} W_{n,m}(\beta)=0$. This is consistent with the lack of zero-threshold  fermionic superradiance in 2D systems in other theoretical works \cite{piazza2014Umklapp, chen2014superradiance} as well as in experiments \cite{helson2023densitywave, zhang2021observation}.

\section{Momentum density deformation and soft-transition superradiance}
In the previous sections, we obtained a quantitative argument on a zero-threshold fermionic superradiant phase transition based on a Pauli crystal degeneracy splitting due to the Hamiltonian term associated with scattering of photons into the cavity. In this section we aim to give a physical intuition behind the analytics previously shown. In particular, we will show that this degeneracy splitting and a non-zero value of $\ev*{\ca} = \alpha$ is associated to an elongation of the Fermi sea along the direction given by either of $\vb{q}_{\pm} = \vb{k}_p \pm \vb{k}_c$, the scattering wavevectors that send photons to the cavity from the transverse pump laser and vice-versa. 

We start from the Hamiltonian from the main text
\begin{equation}
  \Ha = \Ha[0]  - (\Dc - \hat{\mathcal{B}}) \cad\ca + \hat{V}
\end{equation}
where $\Ha[0]$ contains the kinetic energy and trap terms,  $\Dc = \omega_p - \omega_c<0$ is the cavity detuning,  $\hat{\mathcal{B}} = \int \dd \vb{x} \FOdx U_0 \sin^2(\vb{k}_c \vb{x}) \FOx$ is called ``bunching parameter'', which captures the dispersive shift of the cavity resonance, and $\hat{V} \propto \ca +\cad$ is the interaction Hamiltonian from Eq. (\ref{eq:S_Vint}). By taking an expectation value of this Hamiltonian with respect to the atomic ground state [Pauli crystal] we trace out the atomic degrees of freedom and obtain a Hamiltonian for a displaced harmonic oscillator (up to dropped constant terms)
\begin{align}
  \Ha[c]\equiv  & \ev*{\Ha}_{\text{a}}\\
   = & -\wDc \cad \ca +  \sqrt{U_0 V_0}(F_+ - F_-)(\ca+\cad), \label{eq:S_Hc}
\end{align}
where we introduced the effective cavity detuning $\wDc = \Dc - \ev*{\hat{\mathcal{B}}}_{\text{a}}<0$ and total forcing term 
\begin{equation}
  \hat{F}_{\text{tot}}=\int \dd \vb{x} \FOdx \sin(\vb{k}_p \vb{x}) \sin(\vb{k}_c \vb{x}) \FOx, 
\end{equation}
whose expectation value $ \ev*{\hat{F}_{\text{tot}} }_{\text{a}}= {F}_{\text{tot}}= F_+ - F_-$ may lead to a non-zero expectation value of the cavity field $\alpha$, i.e. it leads to zero-threshold superradiance.

We express this last operator in terms of a momentum space representation:
\begin{align}
  \hat{F}_{\text{tot}}= & -\frac{1}{2}\sum_{\gamma\in\{\pm\}}\gamma \int \dd \vb{x} \FOdx  \cos(\vb{q}_\gamma \vb{x}) \FOx\\
  = &  -\frac{1}{4}\sum_{\vb{q}_\gamma\in\{\pm \vb{q}_{\pm}\}}\gamma \int \dd \vb{x} \FOdx  e^{i \vb{q}_\gamma \vb{x}} \FOx \,,
\end{align}
where we first used a trigonometric identity $ \sin A \sin B = -\sum_{\gamma}\cos(A+\gamma B)/2$. Later, we can expand the field operators as  $\FOx = \frac{1}{\sqrt{2\pi}} \int \dd \vb{k} e^{i \vb{kx}} \FO[\vb{k}]  $ to change into  a momentum space expression, where the integral over space will lead to a Dirac delta $\delta(\vb{k}-\vb{k}' + \vb{q}_\gamma)$, resulting in
\begin{align}
  {F}_{\text{tot}}= & -\frac{1}{4}\sum_{s, \gamma\in\{\pm \}}\gamma \int \dd \vb{k} \ev*{\FOd[\vb{k}+ s \vb{q}_\gamma] \FO[\vb{k}]}_{\text{a}}\\
  = & -\frac{1}{4}\sum_{s, \gamma\in\{\pm \}}\gamma \int \dd \vb{k} N \rho(\vb{k}+s \vb{q}_\gamma, \vb{k})\\
  % = & -\frac{\sqrt{U_0 V_0}}{2} \sum_{\gamma\in\{\pm\}}\gamma \int \dd \vb{k} N \rho(\vb{k}+\vb{q}_\gamma, \vb{k})\\
  \equiv & \sum_{\gamma\in\{\pm\}}\gamma F_\gamma,
\end{align}
where we first identified  $\ev*{\FOd[\vb{k}+\vb{q}_\gamma] \FO[\vb{k}]}_{\text{a}}/N = \rho(\vb{k}+\vb{q}_\gamma, \vb{k})$ as the one-body density matrix in momentum space and  defined our forcing terms $F_{+} = F(\vb{q}_+)$ and $F_{-} = F(\vb{q}_-)$: 
\begin{align}
  F_\gamma = &  -\frac{N }{4} \sum_{s\in\{\pm\}} \int \dd \vb{k}  \rho(\vb{k}+s \vb{q}_\gamma, \vb{k})  \label{eq:S_defFgamma}    \\
  = &-\frac{1}{2}  \int \dd \vb{x} \ev*{ \FOdx  \cos(\vb{q}_\gamma \vb{x}) \FOx},  \label{eq:S_defFgx}
\end{align}
where the last expression is an equivalent position space quantity. Notice that the $\int \dd \vb{k} \rho(\vb{k}+ \vb{q}_\gamma, \vb{k})$ from Eq. (\ref{eq:S_defFgamma}) is a geometrical measure of the extent of $\rho(\vb{k})$ along the axis given by $\vb{q}_\gamma$. Moreover, it can be considered as an average correlation of the momentum density at a point in momentum space $\vb{k}_0$ with another one at $\vb{k}_0+\vb{q}_\gamma$.

One can use the expression for $F_\gamma$ from Eq. (\ref{eq:S_defFgx}) together with the definitions in Eq. (\ref{eq:S_ei}) to analytically obtain the values for $F_\pm$. This requires using the fact that for $\sigma = \pm 1 $: $Y_{n,m}(\sigma\beta) = Y_{n,m}(\beta)$ and $W_{n,m}(\sigma\beta) = \sigma W_{n,m}(\beta)$, as well as $\vb{q}_\gamma = \vb{k}_p + \gamma \vb{k}_q = (\gamma \hat{x}+ \hat{y}) \pi \beta/L$.  Using the states that maximize the atomic subspace degeneracy splitting, one obtains:
\begin{align}
  F_\gamma = - \frac{1}{2}  \Bigg[ & \mp \gamma  W_{a,b}^2 + Y_{a,a}Y_{b,b} \pm  Y_{a,b}^2  \nonumber \\ 
  &+\sum_{\underline{n}=1}^{N-1} Y_{n_x,n_x}  Y_{n_y,n_y} \Bigg],
\end{align}
 which directly gives 
\begin{align}
  F_+ - F_-  = \pm  W_{a,b}^2(\beta), \label{eq:S_FpFm}
\end{align}
in agreement with our previous calculation from Eq. (\ref{eq:S_Va}). 

The term ${F}_{\text{tot}} =  F_+ - F_-$, with opposite signs for individual ``forces'' together with the form of the cavity Hamiltonian from Eq. (\ref{eq:S_Hc}) suggests that each scattering wavevector $\vb{q}_\pm$ couples an elongation of the momentum distribution to a phase of the scattered light into the cavity. Notice that the expression from Eq. (\ref{eq:S_defFgamma}) holds for all momentum distributions while the one of Eq. (\ref{eq:S_FpFm}) is specific to the distributions that maximize the total forcing term. Moreover, this physical interpretation sheds light on why some degenerate Pauli crystals can scatter light at zero threshold, while others cannot. Take for example the Pauli crystals  for $N=5$ (Fig. \ref{fig:S_bloch}e) and $N=7$ in a box (Fig. \ref{fig:S_bloch}f). One can see that the momentum space distributions associated to $\ket{\xi}$ and $\ket{\zeta}$ for $N=5$ have equal elongation along the axes given by $\vb{q}_\pm$, while those of $N=7$ are clearly elongated along each of them. Furthermore, this gives background to the dependence of $W_{n,m}(\beta)$ on the parity of $n\pm m$ from Eq. (\ref{eq:S_Wnm}) for even values of $\beta$. Different parity values of $n_{x(y)}$ and $m_{x(y)}$ (resulting in $n_{x(y)} \pm m_{x(y)}$ being odd) makes part of the distribution of the Pauli crystal along the $k_x(k_y)$ direction be an odd function. This results in an elongated distribution in momentum space, whereas  if $n_{x(y)} \pm m_{x(y)}$  is even, the momentum distribution is also even and no elongation is obtained.

\section{Simulations in MCTDH-X}

In the main text, the numerical tool MCTDH-X (Multiconfigurational Time-dependent Hartree method for Indistinguishable particles)~\cite{lode2016multiconfigurational, alon2008multiconfigurational, fasshauer2016multiconfigurational} was used to perform simulations of fermions in a cavity. This open-source code (available at \cite{lode2024ultracold}) has been extensively used to study cavity-boson systems in 1D and 2D~\cite{lin2019superfluid, lin2020pathway} and cavity-fermion systems in 1D~\cite{molignini2022crystal}.
The method considers a Hamiltonian consisting of a one-body potential $V(\vb{x})$ and a two-body interaction $W(\vb{x}, \vb{x}')$, which can be, in general, time-dependent:
\begin{align}
    \Ha & =  \int \dd \vb{x} \FOdx \left( -\frac{1}{2m}\nabla^2 +V(\vb{x})\right) \FOx \nonumber \\ &+ \frac{1}{2}\int \dd \vb{x} \dd \vb{x}'  \FOdx \FOd(\vb{x}') W(\vb{x}, \vb{x}') \FO(\vb{x}') \FOx.
    \label{eq:S_HamiltinanMCTDHX}
\end{align}
MCTDH-X is based on the ansatz that a finite number of (time-dependent) single-particle orbitals can describe the system's many-body state. 

Consider a system with $M$ orbitals and $N$ particles which are distributed over them. We define single orbital populations as  $n_1,\dots, n_M$ such that $N=\sum_{k=1}^{M}n_k$. We can write the ansatz $\ket{\Phi}$ as
\begin{equation}
\ket{\Phi} =  \sum_{\{\vb{n}\}}C_{\vb{n}}(t) \prod_{k=1}^{M} \left[ \frac{\left( \hat{b}_k^\dag (t) \right)^{n_k}}{\sqrt{n_k!}} \right]\ket{\text{vac}}, \label{eq: ansatz}
\end{equation}
with a field operator
\begin{equation}
    \FOx =  \sum_{k=1}^M \phi_k(\vb{x};t)\hat{b}_k  (t),
\end{equation}
where $\vb{n} = (n_1, n_2, \dots, n_M)$ and $\hat{b}_k^\dag (t)$ is a creation operator of a particle in the $k$-th orbital with a wavefunction $\phi_k(\vb{x};t)$. Here, $\hat{b}_k (t)$  encodes the particle statistics of the system, i.e., $\{\hat{b}_k, \hat{b}_j^\dag \}=\delta_{k,j}$ for fermions.   MCTDH-X is a time-dependent variational method~\cite{Kramer1981} to determine both the coefficients  $C_{\vb{n}}$ and the $M$ orbitals $\phi_k(\vb{x})$ in real or imaginary time. It is exact in the limit of infinite orbitals, but in practice, a finite number $M$ of orbitals can be sufficient.

While $N$ bosons can be simulated in the Gross-Pitaevskii limit using $M=1$ orbital, Pauli exclusion principle demands that $M\geq N$, where the lower bound is a Slater determinant state that cannot be used for an interacting system. Furthermore, simulations of even small numbers of fermions in 2D scale rapidly in particle number. Computational complexity scales as $\smqty(M+N+1\\M)$, meaning this limit cannot be reached in practice for many systems. Recently, the code has been updated to use High Performance Computing (HPC) techniques to make feasible 2D simulations as the ones done for the main text. In particular, Message Passing Interface (MPI) allows for parallelization of the computations using domain decomposed orbitals (DDO).

By its nature, MCTDH-X is designed to address the Schrödinger equation for static, non-dissipative systems. However, it is possible to introduce a dissipative cavity as is done in the cavity module of MCTDH-X \cite{lode2017fragmented}. Consider a Lindbladian consisting of the Hamiltonian from Eq. (\ref{eq:S_HamiltinanMCTDHX}), which also includes cavity photon creation (annihilation) operators $\ca$ ($\cad$), and a jump operator $\hat{L} = \kappa \ca$. If we consider the mean-field limit $\ca \rightarrow \alpha = \ev{\ca}$, the particles have an evolution described by the variational principle applied to the MCTDH-X ansatz. Meanwhile, the cavity field will evolve according to 
\begin{align}
  \pdv{t}\alpha = & i \ev{[\Ha, \ca]}_{\text{c}}  -\kappa\alpha, 
\end{align}
where $\ev{\cdot}_{\text{c}}$ takes an expectation value of the cavity operators in the mean-field limit. This equation of motion is solved to find the steady state solution at every time step of the method. 

In the main text, only spin-polarized fermions were considered, i.e. the only interactions between the atoms were induced by the cavity photons. Since the mean-field limit was taken for the cavity field, in the mathematical sense the system was non-interacting, meaning that in principle no more than $M$ orbitals were needed. In the simulations presented in the main text, for each particle number \( N \), we fixed \( NU_0 = 2\pi \times 5.22 \, \text{MHz} \) and calculated \( M = N+1 \) orbitals. The resultiing orbitals were obtained by performing successive relaxations in imaginary time until convergence was achieved for both the cavity field \( \alpha(t) \) and the total energy \( E(t) = \langle \mathcal{H}(t) \rangle \).

\begin{figure}[htb]
  \includegraphics[width=1\columnwidth, trim={0 2.5mm 0 0},clip]{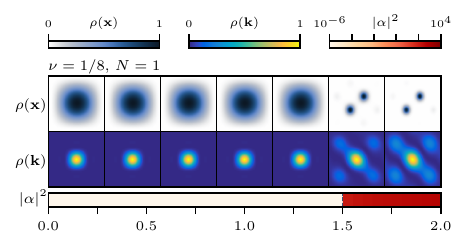}
  \includegraphics[width=1\columnwidth, trim={0 2.5mm 0 0},clip]{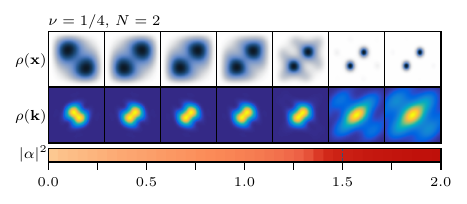}
  \includegraphics[width=1\columnwidth, trim={0 2.5mm 0 0},clip]{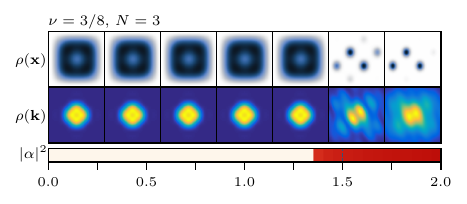}
  \includegraphics[width=1\columnwidth, trim={0 2.5mm 0 0},clip]{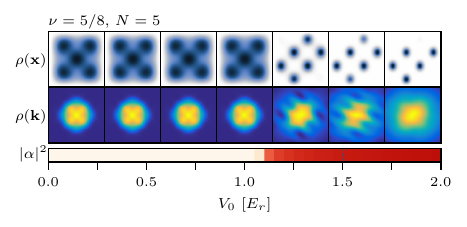}
  \caption{ Pauli crystal superradiance for different fillings $\nu$/ particle numbers $N$. The real ($\rho(\vb{x})$) and momentum space densities ($\rho(\vb{k})$) of Pauli crystals at $N=1,2,3,5$ are shown together with the values of $|\alpha|^2$ in log scale (colorbars) at different values of $V_0$. The densities are shown in $x, y \in [0, 2\lambda]$ and $k_x, k_y \in [-4\pi/\lambda, 4\pi/\lambda]$, and normalized to the corresponding maximum value. 
  }
  \label{fig:S_extra_nu1}
\end{figure} 

\begin{figure}[htb]
  \includegraphics[width=1\columnwidth, trim={0 2.5mm 0 0},clip]{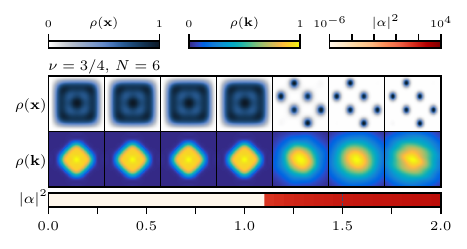}
  \includegraphics[width=1\columnwidth, trim={0 2.5mm 0 0},clip]{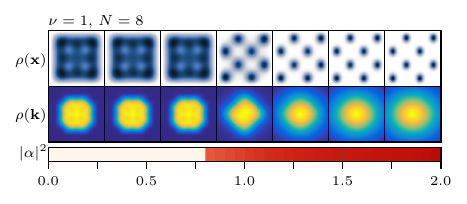}
  \includegraphics[width=1\columnwidth, trim={0 2.5mm 0 0},clip]{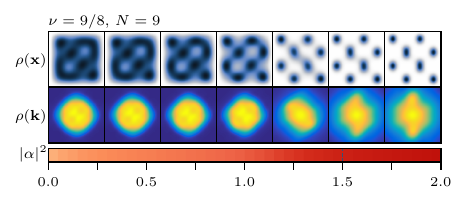}
  \includegraphics[width=1\columnwidth, trim={0 2.mm 0 0},clip]{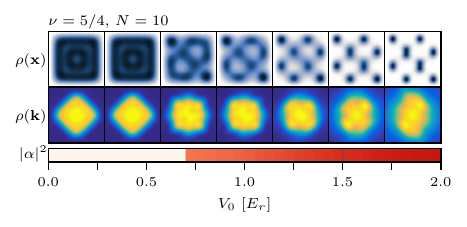}
  \caption{
  Pauli crystal superradiance for different fillings $\nu$/ particle numbers $N$. The real ($\rho(\vb{x})$) and momentum space densities ($\rho(\vb{k})$) of Pauli crystals at $N=6,8,9,10$ are shown together with the values of $|\alpha|^2$ in log scale (colorbars) at different values of $V_0$. The densities are shown in $x, y \in [0, 2\lambda]$ and $k_x, k_y \in [-4\pi/\lambda, 4\pi/\lambda]$, and normalized to the corresponding maximum value.
  }
  \label{fig:S_extra_nu2}
\end{figure}

\begin{figure}[htb]
  \includegraphics[width=1\columnwidth, trim={0 2.5mm 0 0},clip]{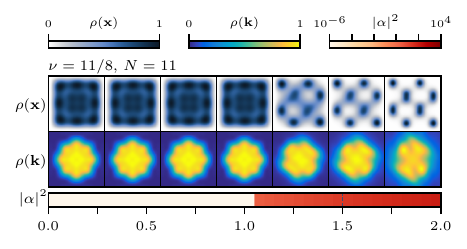}
  \includegraphics[width=1\columnwidth, trim={0 2.5mm 0 0},clip]{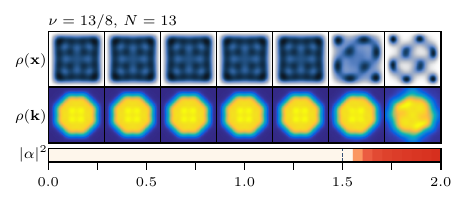}
  \includegraphics[width=1\columnwidth, trim={0 2.5mm 0 0},clip]{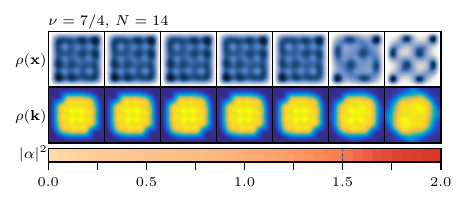}
  \includegraphics[width=1\columnwidth, trim={0 2.5mm 0 0},clip]{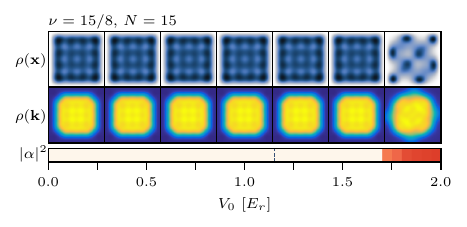}
  \caption{ 
  Pauli crystal superradiance for different fillings $\nu$/ particle numbers $N$. The real ($\rho(\vb{x})$) and momentum space densities ($\rho(\vb{k})$) of Pauli crystals at $N=11,13,14,15$ are shown together with the values of $|\alpha|^2$ in log scale (colorbars) at different values of $V_0$. The densities are shown in $x, y \in [0, 2\lambda]$ and $k_x, k_y \in [-4\pi/\lambda, 4\pi/\lambda]$, and normalized to the corresponding maximum value.
  }
  \label{fig:S_extra_nu3}
\end{figure}

\section{Superradiance transitions for other filling fractions $\nu$}

In this section, we present the simulations for $N=1,2,...,15$ particles which were not included in the main text. Ground states for $N=1,2,3,5$ can be found in Fig. \ref{fig:S_extra_nu1}, $N=6,8,9,10$ in Fig. \ref{fig:S_extra_nu2} and $N=11,13,14,15$ in Fig. \ref{fig:S_extra_nu3}. Soft transitions are present only in $N=2, 9$ and $14$.

\begin{figure}[htb]
  \includegraphics[width=1\columnwidth]{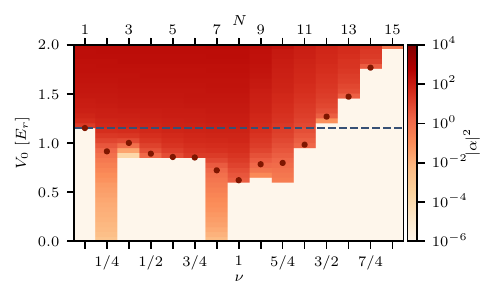} 
  \vspace{-7mm}
  \caption{ Phase diagram for a system including a transverse pump along the $y$ direction. Soft-transitions are only attained for $N=2$ and $7$ fermions, while the other soft-transitions obtained in the main text at $N=9$ and $14$ are suppressed. The cavity field (colorbar in log-scale) is shown as a function of pump strength $V_0$ for Pauli crystals with different filling factors $\nu$. Zero threshold transitions take place when the most energetic fermion of an open-shell Pauli crystal is susceptible to the cavity for $\nu<1$. As a visual aid and for comparison, dots mark the (arbitrary) $|\alpha|^2 = 1$ threshold, while the dashed line indicates the bosonic threshold. }  
  \label{fig:figS_phaseDiag_wTP}
\end{figure}

\begin{figure}[htb]
  % \vspace{-5mm}
  \includegraphics[width=1\columnwidth, trim={0 2.5mm 0 0},clip]{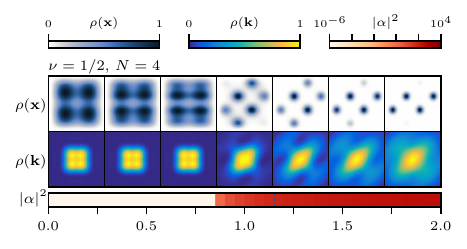}
  \includegraphics[width=1\columnwidth, trim={0 2.5mm 0 0},clip]{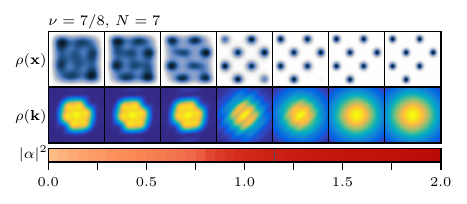}
  \includegraphics[width=1\columnwidth, trim={0 2.5mm 0 0},clip]{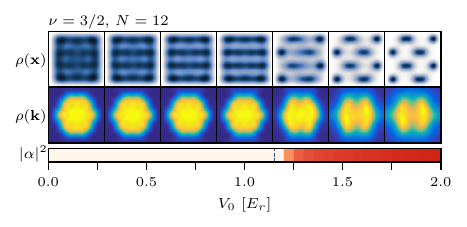}
  \caption{ Pauli crystal superradiance for different fillings $\nu$/ particle numbers $N$ in the case of an included transverse pump lattice $\hat{V}_p$ in the $y$ direction. The real ($\rho(\vb{x})$) and momentum space densities ($\rho(\vb{k})$) of Pauli crystals at $N=4,7,9,12$ are shown together with the values of $|\alpha|^2$ in log scale (colorbars) at different values of $V_0$. The densities are shown in $x, y \in [0, 2\lambda]$ and $k_x, k_y \in [-4\pi/\lambda, 4\pi/\lambda]$, and normalized to the corresponding maximum value. 
  }
  \label{fig:S_wTP}
\end{figure} 

\newpage
\section{Effect of a transverse pump standing wave in soft transition}

Typical experimental realizations of our cavity-atom model can contain additional contributions such as a transverse pump standing wave optical lattice potential $\hat{V}_{p}=\int \dd \vb{x} \FOdx V_0 \sin^2(\vb{k}_p\vb{x}) \FOx$ added to the Hamiltonian~\cite{baumann_2010_dicke}. As opposed to the main text Hamiltonian $\Ha$ [cf. Eq.(1)], which stays as the bare kinetic energy and trap terms included in $\Ha[0]$ before there is any light in the cavity for any value of $V_0$, the addition of $\hat{V}_{p}$ adds a modulation along the $y$ axis in position space for $\rho(\vb{x})$ and sidebands around $\vb{k} = \pm 2\vb{k}_p$ showing momentum recoil processes in $\rho(\vb{k})$. Here, we show that our results in the main text are robust against this kind of change. In particular, for filling fractions below $\nu=1$ we have a qualitatively similar behavior as we showed in the main text. However, for $\nu >1$ we have that the transverse pump standing wave cancels the soft-transitions. As we will show, this is due to the lattice term breaking the degeneracy of the degenerate Pauli crystals  $\ket{1}$ and $\ket{2}$. 

Let's consider an interaction Hamiltonian $\Ha[\text{int}] = \hat{V}+ \hat{V}_{p}$, where $\hat{V}$ is the interaction term defined in Eq. (\ref{eq:S_Vint}). Within the degenerate subspace of $\Ha[0]$ we have
\begin{align}
    \Ha[\text{int}] =& \mqty(\mel*{1}{ \Ha[\text{int}]}{1} & \mel*{1}{\Ha[\text{int}]}{2} \\
  \mel*{2}{\Ha[\text{int}]}{1} &\mel*{2}{\Ha[\text{int}]}{2}) \label{eq:S_Hint_mat} \\
  = &  \sum_{n=1}^{N-1} \ev*{\Ha[\text{int}]^{(1)}}{\underline{n}}   \mathbb{1} + \nonumber \\ &
  \mqty(\mel*{\underline{d1}}{ \Ha[\text{int}]^{(1)}}{\underline{d1}} & \mel*{\underline{d1}}{\Ha[\text{int}]^{(1)}}{\underline{d2}} \\
  \mel*{\underline{d2}}{\Ha[\text{int}]^{(1)}}{\underline{d1}} &\mel*{\underline{d2}}{\Ha[\text{int}]^{(1)}}{\underline{d1}}) 
\end{align}
where $\mathbb{1}$ is an identity matrix in this degenerate subspace, $\Ha[\text{int}]^{(1)}$ is a one-particle interacting Hamiltonian, $\ket{\underline{n}} = \ket{n_x,n_y}$ is a single particle eigenstate of the bare Hamiltonian with a square box trap and $\ket{\underline{d1}} = \ket{a,b} $ and  $\ket{\underline{d2}} = \ket{b,a} $ are two degenerate eigenmodes of an open-shell. In order to calculate it, we write $\Ha[\text{int}]^{(1)} = \sqrt{U_0 V_0}(\ca+\cad) \sin(2\pi\hat{x}/\lambda) \sin(2\pi\hat{y}/\lambda) + V_0 (1-\cos(4\pi \hat{y}/\lambda)/2$  and use the previous 1D single-particle definitions $W_{n_\mu,m_\mu}(\beta) = \bra{n_\mu}\sin(2 \pi \hat{\mu}/\lambda)\ket{m_\mu}$ and $Y_{n_\mu,m_\mu}(\beta) = \bra{n_\mu}\cos(2 \pi \hat{\mu}/\lambda)\ket{m_\mu}$ with $ \mu \in \{x,y\}$, evaluated in Eqs. (\ref{eq:S_Wnm}-\ref{eq:S_Ynm}). With this we obtain
\begin{align}
    \Ha[\text{int}] =& \frac{ V_0}{2}\left[ N - \sum_{n=1}^{N-1} Y_{n_y, n_y}(2\beta) \right] \mathbb{1} \nonumber\\ 
    & -\frac{V_0}{2} \mqty(Y_{b,b}(2\beta) & 0 \\
   0 &  Y_{a,a}(2\beta)) \nonumber \\
    &+ \sqrt{U_0 V_0}(\ca+\cad)W_{a,b}^2(\beta) \mqty(0 & 1  \\
  1 & 0) \\
  = & \frac{ V_0}{2}\left[ N + \sum_{n=1}^{N-1} \delta_{n_y, \beta} \right] \mathbb{1} +\frac{V_0}{4} \mqty(\delta_{b, \beta} & 0 \\
   0 &  \delta_{a, \beta}) \nonumber \\
    &+ \sqrt{U_0 V_0}(\ca+\cad)W_{a,b}^2(\beta) \mqty(0 & 1  \\
  1 & 0),
\end{align}
where  the terms with $\delta_{a, \beta}$ and $\delta_{b, \beta}$ can lift the degeneracy when $a=\beta$ or $b=\beta$. This occurs at $\nu=1$ or at the next immediately larger filling fraction, depending on the value of $\beta$. Thus, at $\nu =1$ our degenerate perturbation theory argument no longer holds and there is a single Pauli crystal configuration which is the ground state of the system. Then, there is no possibility of using a degenerate atomic state $\ket{\pm}_a$ which can lead to a non-zero order parameter, thus removing the soft-transition. In practice, we observed in our simulations that for all particle numbers above $\nu=1$ soft transitions no longer appeared. This means that, in a similar fashion, degeneracy is lifted after $\nu=1$.

Here, we present the simulation results for $N=1, 2, \dots, 15$ in the case where the Hamiltonian included a transverse pump lattice $\hat{V}_p = \int\dd \vb{x} \FOdx V_0 \sin^2(\vb{k}_p \vb{x})\FOx$. We present the phase diagram for this Hamiltonian in Fig. \ref{fig:figS_phaseDiag_wTP} and the corresponding ground states for $N=4,7, 12$ can be found in Fig. \ref{fig:S_wTP}.

% \bibliography{apssamp}

\end{document}